\documentclass[12pt]{article}
\usepackage{epsf}
\usepackage{graphicx}

\begin{document}

\title{New Regular Solutions
with Axial Symmetry
in Einstein-Yang-Mills Theory}

\vspace{1.5truecm}
\author{
{\bf Rustam Ibadov}\\
Department of Theoretical Physics and Computer Science,\\
Samarkand State University, Samarkand, Usbekistan\\
and\\
{\bf Burkhard Kleihaus, Jutta Kunz and Yasha Shnir}\\
Institut f\"ur Physik, Universit\"at Oldenburg, Postfach 2503\\
D-26111 Oldenburg, Germany}

\vspace{1.5truecm}

\maketitle
\vspace{1.0truecm}

\begin{abstract}
We construct new regular solutions in Einstein-Yang-Mills theory.
They are static, axially symmetric and asymptotically flat.
They are characterized by a pair of integers $(k,n)$,
where $k$ is related to the polar angle and $n$ to the 
azimuthal angle.
The known spherically and axially symmetric EYM solutions
have $k=1$.
For $k>1$ new solutions arise, which form two branches.
They exist above a minimal value of $n$, that increases with $k$.
The solutions on the lower mass branch are related to
certain solutions of Einstein-Yang-Mills-Higgs theory,
where the nodes of the Higgs field form rings.
\end{abstract}
\vfill\eject

\section{Introduction}

The well-known regular Bartnik-McKinnon (BM) solutions \cite{bm}
and the corresponding non-Abelian black hole solutions \cite{bh},
are asymptotically flat, static spherically symmetric solutions
of SU(2) Einstein-Yang-Mills (EYM) theory.
They are unstable solutions, sphalerons \cite{stab}, and are
characterized by the number of nodes of the gauge field.
Besides the BM solutions
there are also asymptotically flat, static, only axially symmetric 
regular and black hole solutions \cite{kk}.
These are characterized by two integers,
the node number of their gauge field function(s),
and the winding number 
with respect to the azimuthal angle, denoted $n$.
The spherically symmetric solutions have winding number $n=1$,
while the axially symmetric solutions have winding number $n>1$.

In SU(2) Einstein-Yang-Mills-Higgs (EYMH) theory, 
with a triplet Higgs field,
gravitating monopole solutions arise \cite{mono}.
While gravitating monopoles with unit magnetic charge
are spherically symmetric, the known gravitating multimonopoles possess
axial symmetry \cite{hkk}. 
As in EYM theory, these EYMH solutions are characterized by two integers,
the node number of the gauge field and the azimuthal winding number $n$,
which corresponds to the topological charge of the monopoles.

But EYMH theory allows for further static axially symmetric solutions,
representing gravitating monopole-antimonopole pair, chain and vortex
solutions \cite{map,kks,kksi}.
These solutions can be characterized by the azimuthal winding number $n$,
and by a second integer $m$,
related to the polar angle.
For the monopole-antimonopole chains, 
which in flat space arise for $n=1$ and 2, 
the integer $m$ corresponds to the number of nodes of the Higgs field
(and thus the number of poles on the symmetry axis) \cite{kks}.
In vortex solutions, on the other hand, 
which in flat space arise for winding number $n>2$,
the Higgs field vanishes (for even $m$)
on $m/2$ rings centered around the symmetry axis \cite{kks}.

The existence of monopole-antimonopole pair, 
chain and vortex solutions in EYMH theory \cite{map,kks,kksi}
immediately leads to the question, whether there might be
analogous solutions in EYM theory, even though there would be no
Higgs field participating in the subtle interplay of attraction and
repulsion. 
In this letter we report the existence of one such new type of solution,
related to vortex solutions in EYMH theory.

In section II we present the EYM action, the axially
symmetric ansatz and the boundary conditions. 
In section III we discuss
the properties of the new axially symmetric solutions,
and we present our conclusions in section IV.

\section{Action and Ansatz}

We consider the SU(2) EYM action
\begin{equation}
S=\int \left ( \frac{R}{16\pi G} 
-\frac{1}{2} {\rm Tr} (F_{\mu\nu} F^{\mu\nu}) \right ) \sqrt{-g} d^4x
\ \label{action} \end{equation}
with Ricci scalar $R$,
field strength tensor
\begin{equation}
F_{\mu \nu} = 
\partial_\mu A_\nu -\partial_\nu A_\mu + i e \left[A_\mu , A_\nu \right] 
\ , \label{fmn} \end{equation}
gauge potential $ A_{\mu} = \tau^a A_\mu^a/2 $,
and gravitational and Yang-Mills coupling constants $G$ and $e$,
respectively.
Variation of the action (\ref{action}) with respect to the metric
$g^{\mu\nu}$ leads to the Einstein equations,
variation with respect to the gauge potential $A_\mu$ 
to the gauge field equations.

In isotropic coordinates
the static axially symmetric metric reads \cite{kk}
\begin{equation}
ds^2=
  - f dt^2 +  \frac{m}{f} d r^2 + \frac{m r^2}{f} d \theta^2 
           +  \frac{l r^2 \sin^2 \theta}{f} d\varphi^2
\ , \label{metric} \end{equation}
where the metric functions
$f$, $m$ and $l$ are functions of 
the coordinates $r$ and $\theta$, only.
The $z$-axis ($\theta=0$, $\pi$) represents the symmetry axis.
Regularity on the $z$-axis requires $m=l$ there.

For the gauge field we employ the ansatz \cite{kk,kks,kksi}
\begin{equation}
A_\mu dx^\mu =
\frac{1}{2er} \left[ \tau^n_\varphi 
 \left( H_1 dr + \left(1-H_2\right) r d\theta \right)
 -n \left( \tau^{n,k}_r H_3 + \tau^{n,k}_\theta H_4 \right)
  r \sin \theta d\phi \right]
\ . \label{gf1} \end{equation}
Here the symbols $\tau^{n,k}_r$, $\tau^{n,k}_\theta$ and $\tau^n_\varphi$
denote the dot products of the cartesian vector
of Pauli matrices, $\vec \tau = ( \tau_x, \tau_y, \tau_z) $,
with the spatial unit vectors
\begin{eqnarray}
\vec e_r^{\ n,k}      &=& 
(\sin k \theta \cos n \varphi, \sin k \theta \sin n \varphi, \cos k\theta)
\ , \nonumber \\
\vec e_\theta^{\ n,k} &=& 
(\cos k \theta \cos n \varphi, \cos k \theta \sin n \varphi,-\sin k \theta)
\ , \nonumber \\
\vec e_\varphi^{\ n}   &=& (-\sin n \varphi, \cos n \varphi,0) 
\ , \label{rtp} \end{eqnarray}
respectively.
The gauge field functions $H_i$, $i=1-4$, depend on
the coordinates $r$ and $\theta$, only.
For $k=n=1$ and $H_1=H_3=0$, $H_2=1-H_4=w(r)$
the BM solutions \cite{bm} are recovered,
while for $k=1$, $n>1$, one obtains 
the axially symmetric solutions of \cite{kk}.
The new solutions reported here are obtained for $k>1$.
They are related to EYMH solutions with $m=2k$ in the limit
of vanishing Higgs field \cite{kksi}.

The ansatz is form-invariant under the abelian gauge transformation
\cite{kk}
\begin{equation}
 U= \exp \left({\frac{i}{2} \tau^n_\phi \Gamma(r,\theta)} \right)
\ .\label{gauge} \end{equation}
We fix the gauge by choosing the gauge condition \cite{kk,kks,kksi}
\begin{equation}
 r \partial_r H_1 - \partial_\theta H_2 = 0 
\ . \label{gc1} \end{equation}

To obtain asymptotically flat solutions
which are globally regular
and possess the proper symmetries,
we need to impose appropriate boundary conditions \cite{kk,kks,kksi}.
At the origin we impose the boundary conditions
\begin{equation}
\partial_r f = \partial_r m = \partial_r l = 0\ , \ \ \
H_1=H_3=H_4=0\ , \  H_2=1\ ,
\end{equation}
at infinity we impose 
\begin{equation}
f = m = l = 1 \ , \ \ \
H_1 =H_3=0 \ , \ H_2 = 1 - 2k \ , \ 
H_4 =2 \sin(k\theta)/\sin\theta \ ,
\label{bc} \end{equation}
and on the $z$-axis we impose
\begin{equation}
\partial_\theta f=\partial_\theta m=\partial_\theta l =0 \ , \ \ \
 H_1=H_3=0\ , \ \partial_\theta H_2=\partial_\theta H_4=0 \ .
\end{equation}

We further introduce 
the dimensionless coordinate $x$, and the dimensionless mass $\mu$,
\begin{equation}
x=\frac{e}{\sqrt{4\pi G}} r \ , \ \ \
\mu = \frac{e G}{\sqrt{4 \pi G }} M
\ . \label{dimx} \end{equation}

\section{\bf Numerical Results}

Subject to the above boundary conditions,
we solve the system of seven coupled non-linear partial
differential equations numerically.
To map spatial infinity to the finite value $\bar{x}=1$,
we employ the radial coordinate \cite{kk,kks,kksi}
\begin{equation}
\bar{x} = \frac{x}{1+x}
\ . \label{barx} \end{equation}
The numerical calculations are based on the Newton-Raphson method,
and are performed with help of the program FIDISOL \cite{schoen}.
The equations are discretized on a non-equidistant
grid in $\bar{x}$ and  $\theta$.
Typical grids used have sizes $70 \times 30$, 
covering the integration region 
$0\leq\bar{x}\leq 1$ and $0\leq\theta\leq\pi/2$.
For the method, it is essential to have a good
first guess, to start the iteration procedure.
For the $k=1$ EYM solutions, the $n=1$ BM solutions serve as a
first guess, and then the `parameter' $n$ is varied (via unphysical
noninteger values) to obtain solutions with
higher winding number $n$ \cite{kk}.
For the $k=2$ solutions we employ the
$m=4,n=4$ EYMH vortex solution
with (almost) vanishing Higgs field
as a first guess \cite{kksi},
and then again vary $n$.
Similarly, for the $k=3$ solutions we 
start from the $m=6,n=6$ EYMH vortex solution.

In EYMH theory, the monopole-antimonopole pair (MAP) solution
is obtained, when $m=2k=2$, $n=1$. Here a monopole and an antimonopole
are located symmetrically on the $z$-axis. Similarly,
one might try to find a sphaleron-antisphaleron pair (SAP) solution
in EYM theory. The boundary conditions of such a SAP solution,
however, do not differ from those of a BM solution.
Consequently, when the EYMH coupling constant $\alpha=\sqrt{4 \pi G} v$
(where $v$ is the Higgs field expectation value)
is varied, the gravitating MAP solution
starts from the flat space solution,
reaches a critical solution at a maximal value of $\alpha$,
and then with decreasing $\alpha$ smoothly reaches
the BM solution in the limit $\alpha \rightarrow 0$ (after rescaling
of the radial coordinate) \cite{map}.
Consequently, we do not find a SAP solution in EYM theory,
and similarly, we do not find new solutions, when $m=2k=2$, $n>1$,
since their boundary conditions agree with those of the
axially symmetric solutions of \cite{kk}.

When $m=2k=4$, however, the boundary conditions eq.~(\ref{bc})
for the gauge field
differ from those of the known solutions \cite{bm,kk}.
We therefore may expect the existence of new EYM solutions,
subject to these boundary conditions.
Indeed, for $m=4$, $n=4$, for instance,
when $\alpha$ is varied again, 
the gravitating EYMH vortex solution 
starts from the flat space solution \cite{kks},
reaches a critical solution at a maximal value of $\alpha$,
and then with decreasing $\alpha$ smoothly reaches
a new EYM solution in the limit $\alpha \rightarrow 0$ 
(after rescaling of the radial coordinate) \cite{kksi}.

Let us in the following refer to the solutions
characterized by the integers $k$ and $n$ as $(k,n)$ solutions.
In Fig.~1a we exhibit the energy density of the new 
$(2,4)$ EYM solution.
Its energy density has a pronounced maximum on the $\rho$-axis,
thus the energy density is torus-like. 
The energy density of this $(2,4)$ solution is thus similar to 
the energy density of the 
(lowest mass) known $(1,4)$ EYM solution \cite{kk},
exhibited in Fig.~1b.
But the maximum of the $(2,4)$ solution is smaller than
the maximum of the $(1,4)$ solution, and it is shifted
slightly inwards. 
In addition, the $(2,4)$ solution has a small
saddle point on the $\rho$-axis, located further outwards.
On the other hand
comparison with the corresponding EYMH solution \cite{kks} shows, 
that the energy density
of the EYM solution differs significantly from the energy density of the
flat space $(2,4)$ EYMH solution, whose energy density exhibits
two tori, located symmetrically with respect to the $xy$-plane \cite{kks}.

\noindent
\parbox{\textwidth}{
\vspace{-1.cm}
\centerline{(a)
\epsfxsize=10.cm\epsffile{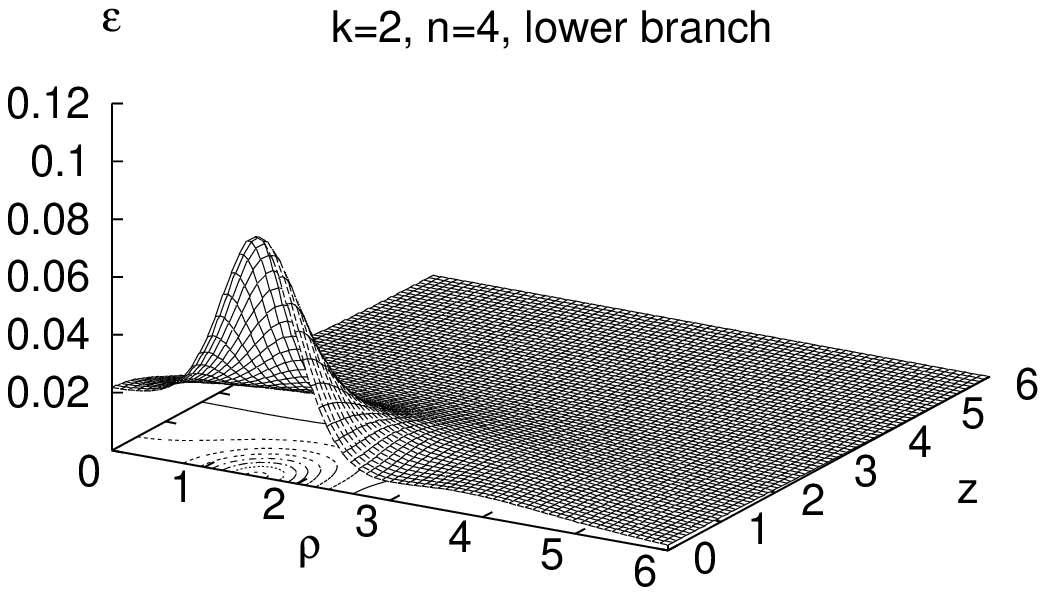}
}
\vspace{-1.cm}
\centerline{(b)
\epsfxsize=10.cm\epsffile{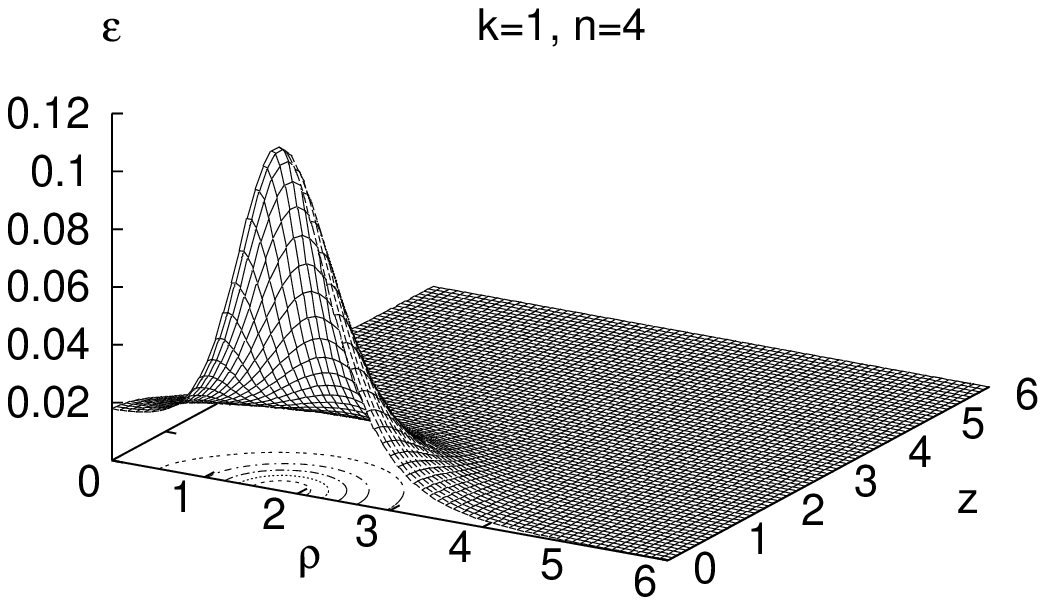} \vspace{0.7cm}
}
{\bf Fig.~1} {\small
The energy density of the $(2,4)$ EYM solution
on the lower mass branch (a) 
and of the $(1,4)$ EYM solution (b).
}\vspace{0.7cm}}

The existence of a $k=2$ EYM solution with winding number $n=4$, immediately 
suggests that analogous solutions with winding number $n \ne 4$ should
exist as well.
Indeed, by increasing $n$, a whole set of such new solutions is obtained
numerically, one for each integer $n$.
On the other hand, a decrease of $n$ does not lead to new solutions.
Instead a bifurcation at the non-integer and thus unphysical value
of $n=3.7175$ is obtained, where a second branch of solutions 
with higher mass appears. 
Consequently there is a second solution for $n=4$, 
and also for each higher integer $n$,
but there are no such solutions for integer $n \le 3$.
This is illustrated in Fig.~2a, where we exhibit
the mass of these $(2,n)$ EYM solutions. 
Only the integer values of $n$ in the figure
correspond to physical solutions.
In Fig.~2b we illustrate 
the value of the metric function $f$ at the origin.
We note, that for $(1,n)$ solutions (with a single node)
we find only one branch, which is also exhibited in Figs.~2.

\noindent
\parbox{\textwidth}{
\centerline{ 
\hspace{0cm}
(a)\epsfxsize=6.cm\epsffile{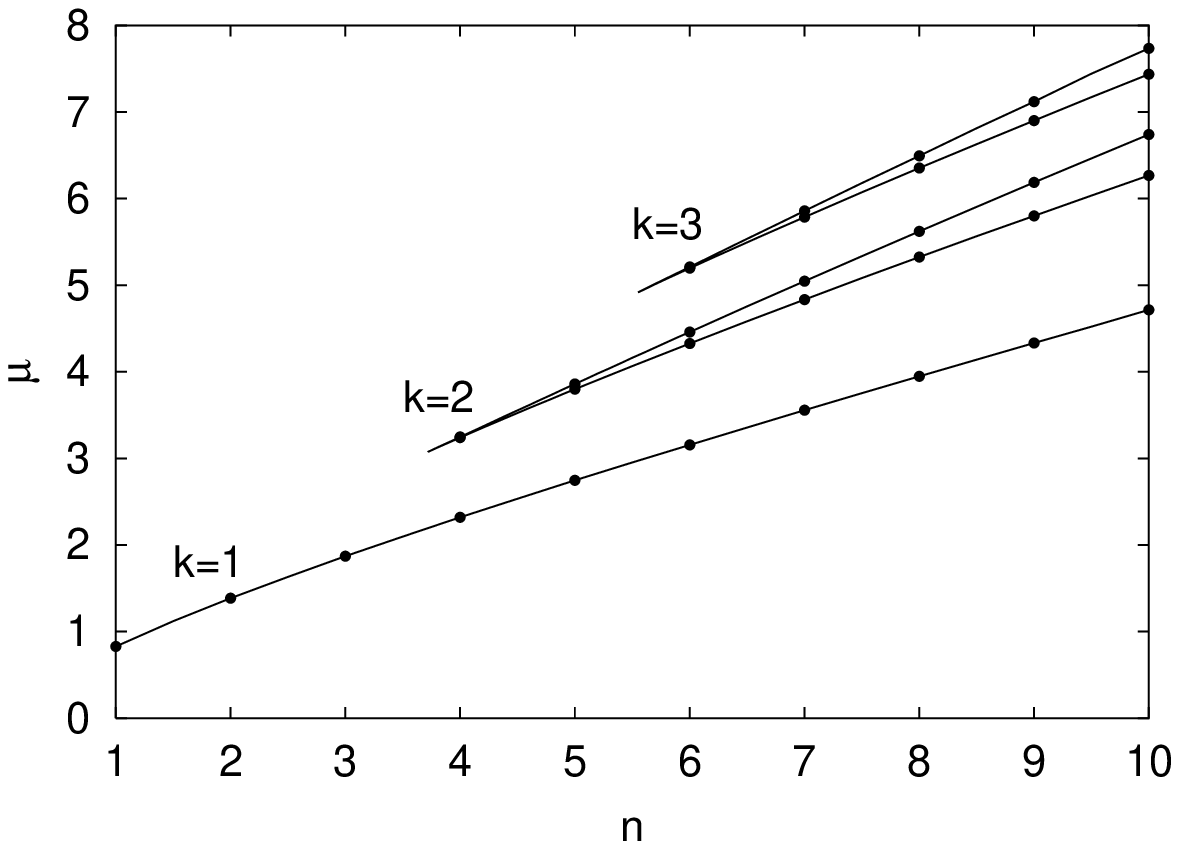}
\hspace{-0cm}(b)
\epsfxsize=6.cm\epsffile{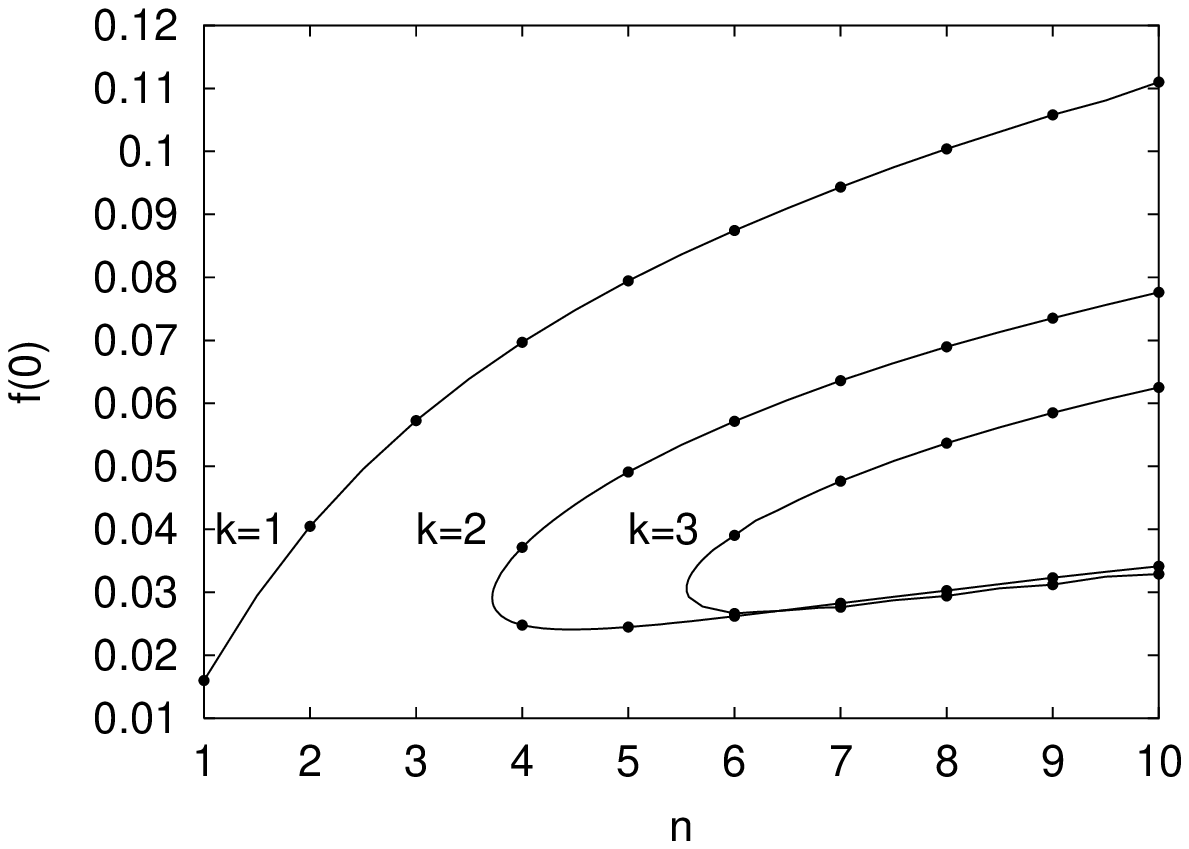} \vspace{0.7cm}
}
{\bf Fig.~2} 
{\small
The mass (a) and the value of the metric function $f$ at the origin (b)
of the $(1,n)$, $(2,n)$ and $(3,n)$ EYM solutions
is shown as a function of $n$.
}\vspace{0.7cm}}

To illustrate the dependence of the solutions on $n$,
we exhibit the energy density of the $(2,n)$ solutions 
with $n=4,6,8$ on the lower (mass) branch in Fig.~3a 
and on the upper branch in Fig.~3b.
With increasing $n$ 
the maximum of the energy density decreases and moves outwards
for the solutions on both branches.
For the same $n$, the maximum of the upper mass solutions
is higher and located further inwards than the maximum
of the lower mass solutions. 
The small saddle point turns into a small maximum on the
upper branch, 
thus the energy density of the solutions 
on the upper branch has a more complicated structure as seen in Fig.~4,
where a surface of constant energy density is shown for the $(2,4)$
solution.
We exhibit the gauge field function $H_2$ for the same set of solutions
in Figs.~5.

\noindent
\parbox{\textwidth}{
\centerline{\hspace{0cm}(a) 
\epsfxsize=6.cm\epsffile{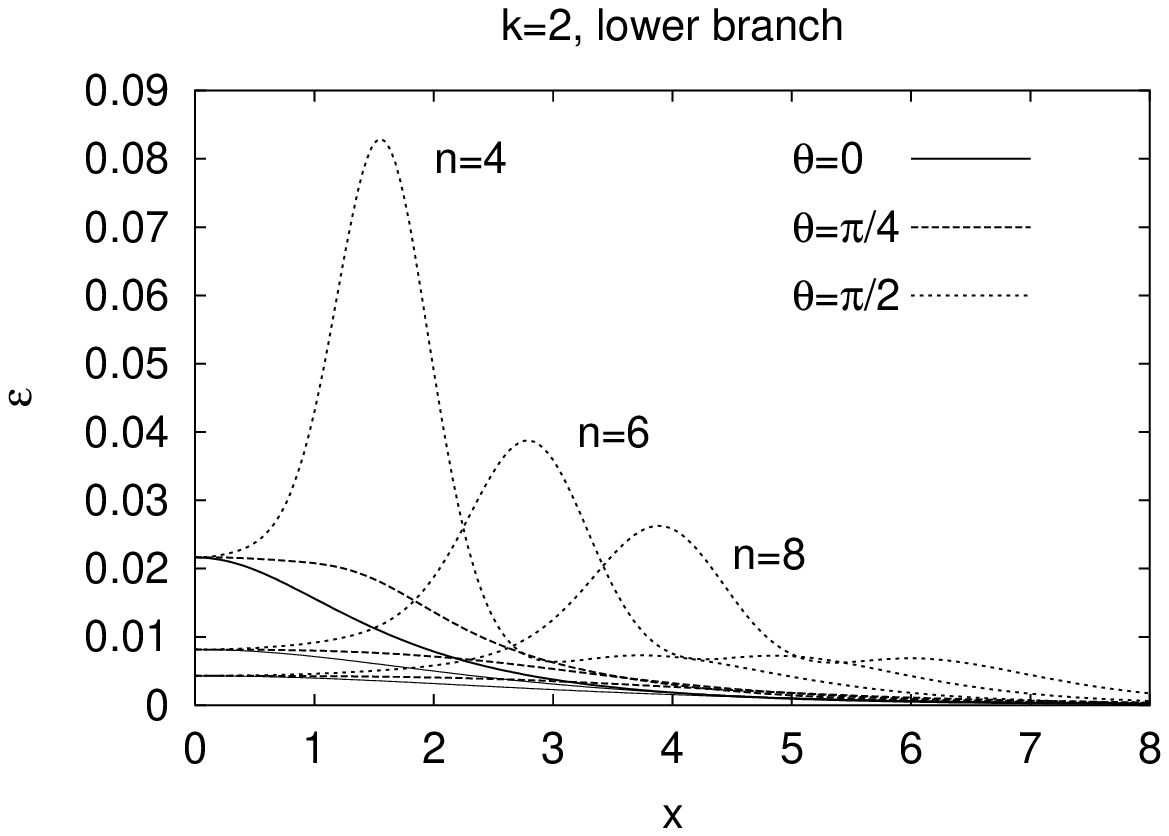}
(b) 
\epsfxsize=6.cm\epsffile{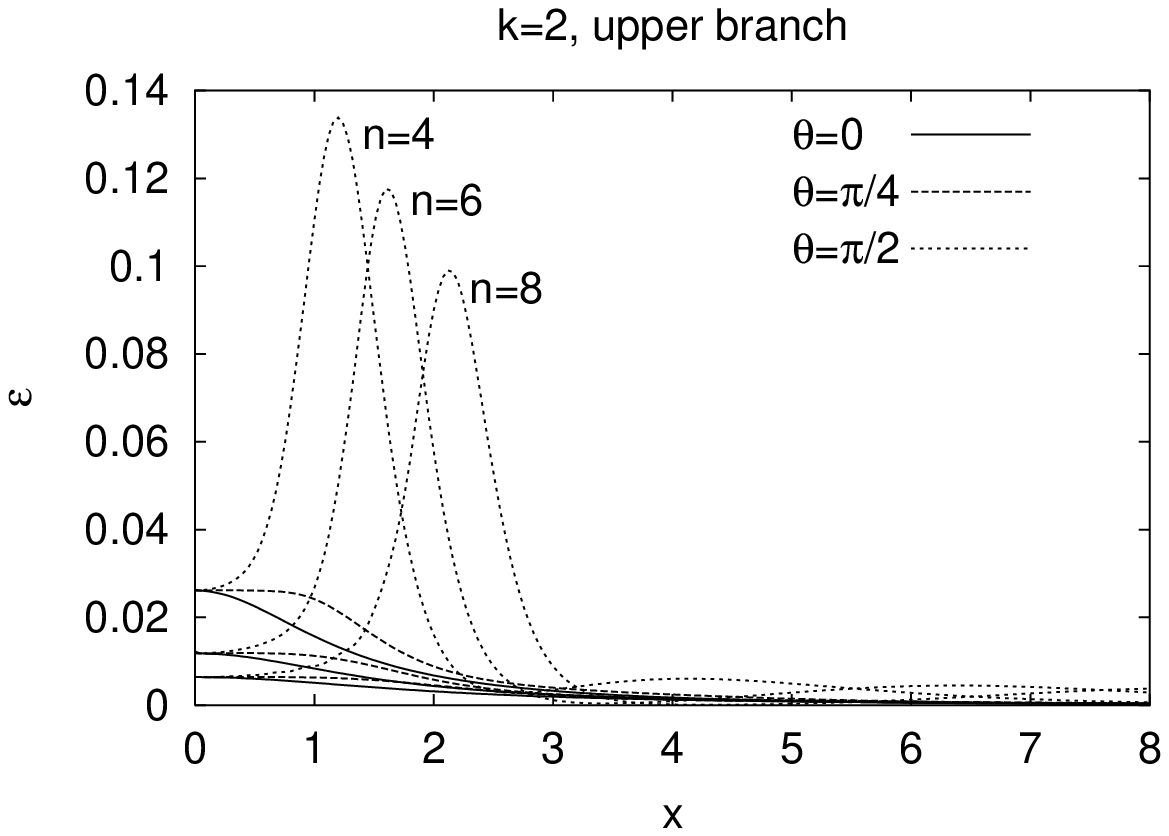}\vspace{0.3cm}
}
{\bf Fig.~3} 
{\small
The energy density of the $(2,n)$ EYM solution on the lower mass branch
(a) and on the upper mass branch (b) for $n=4,6,8$.
}\vspace{0.7cm}}

\noindent
\parbox{\textwidth}{
\centerline{
{\epsfxsize=7.cm\epsffile{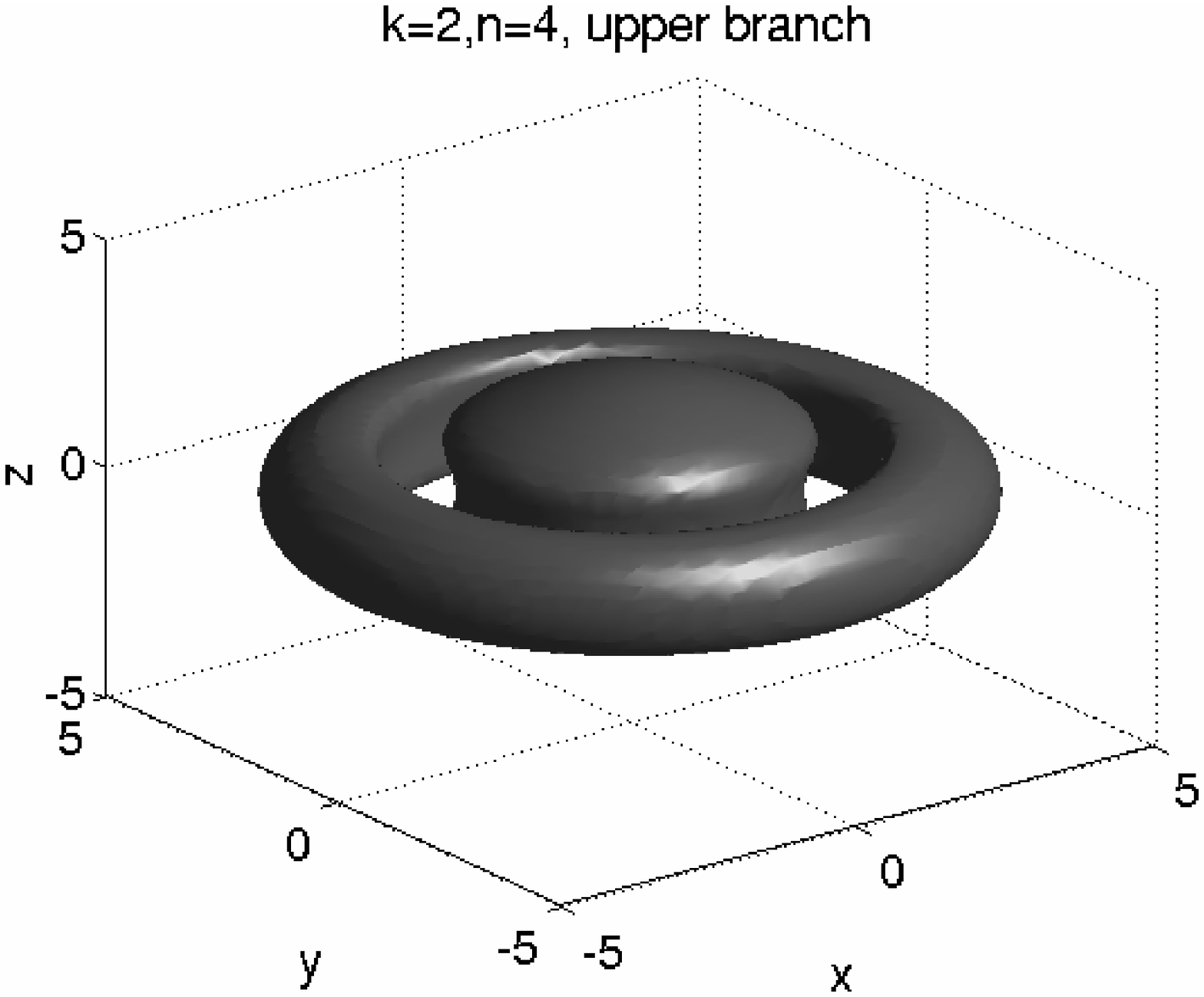}}
}
{\bf Fig.~4}  
{\small
A surface of constant energy density 
(\protect$\varepsilon=0.005$)
of the $(2,4)$ EYM solution on the upper mass branch.
}
\vspace{0.7cm}}

For $k=3$ we obtain a similar pattern of solutions.
But the new type of solutions appears only for $n>5$.
The mass and the value of the metric function $f$ at the origin
of these $(3,n)$ solutions are also shown in Figs.~2.
The energy density 
for $(3,n)$ solutions with $n=6,8,10$ and the
gauge field function $H_2$ are shown in Figs.~6 and 7, respectively.
Based on these results,
we conjecture the existence of $(k,n)$ solutions
with $k>3$, but expect that they will appear
only for still larger values of $n$.

\noindent
\parbox{\textwidth}{
\centerline{(a) 
\epsfxsize=6.cm\epsffile{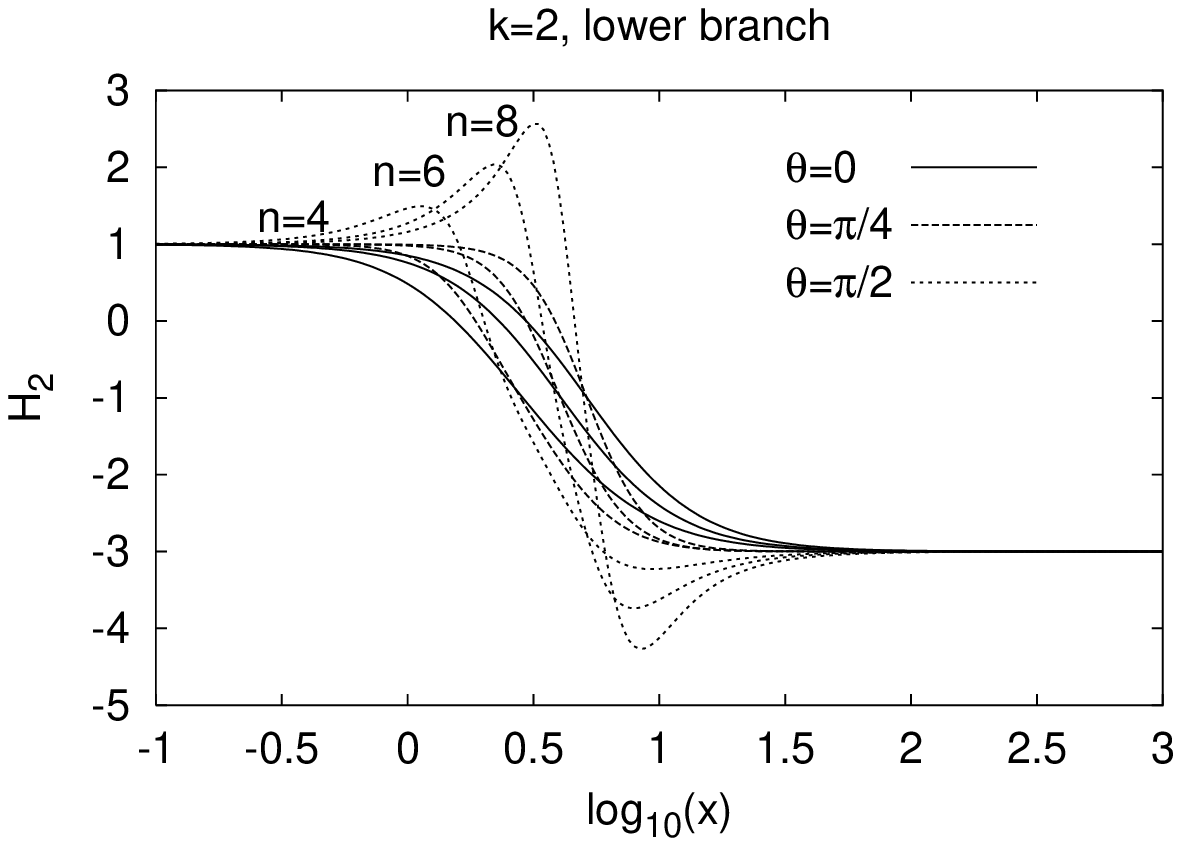}
(b) 
\epsfxsize=6.cm\epsffile{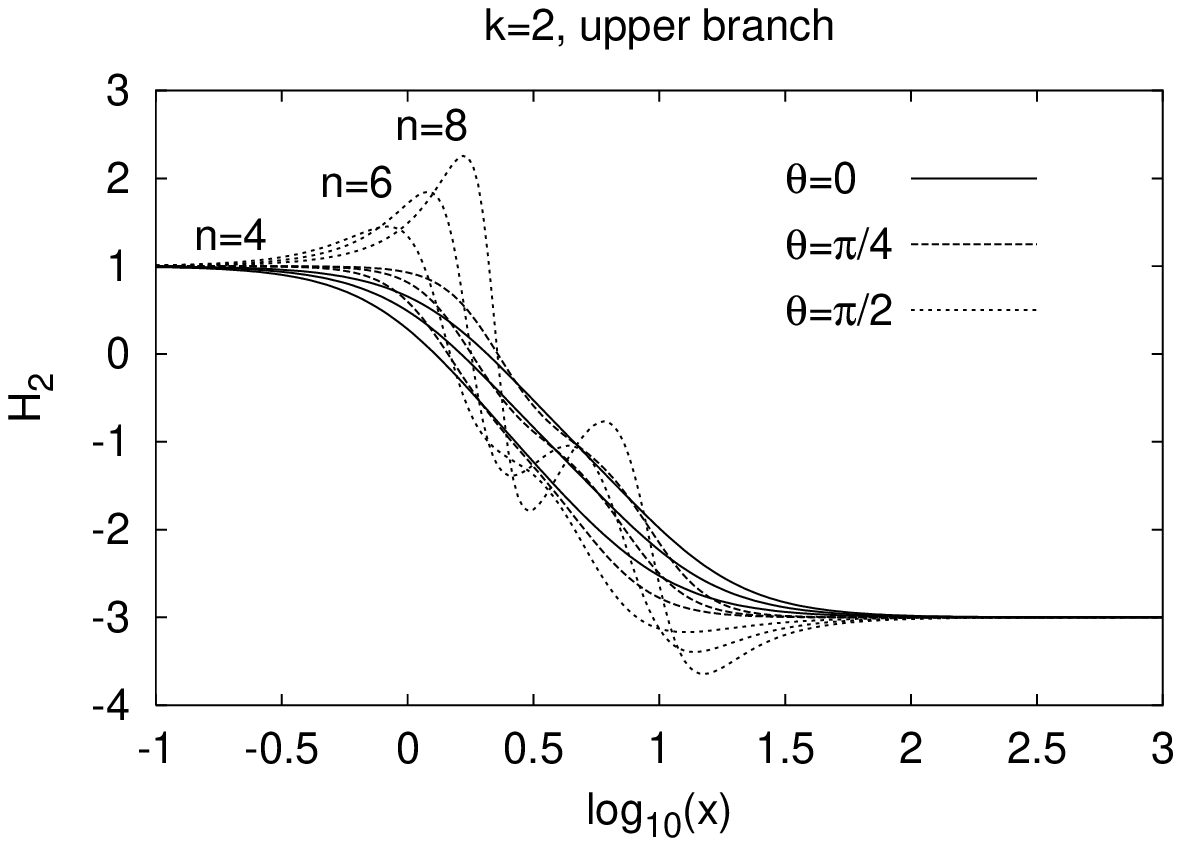}\vspace{0.3cm}
}
{\bf Fig.~5} 
{\small
The gauge field function $H_2$ of the $(2,n)$ EYM solutions
on the lower mass branch (a) 
and on the upper mass branch (b) for $n=4,6,8$.
}\vspace{0.7cm}}

\noindent
\parbox{\textwidth}{
\centerline{(a) 
\epsfxsize=6.cm\epsffile{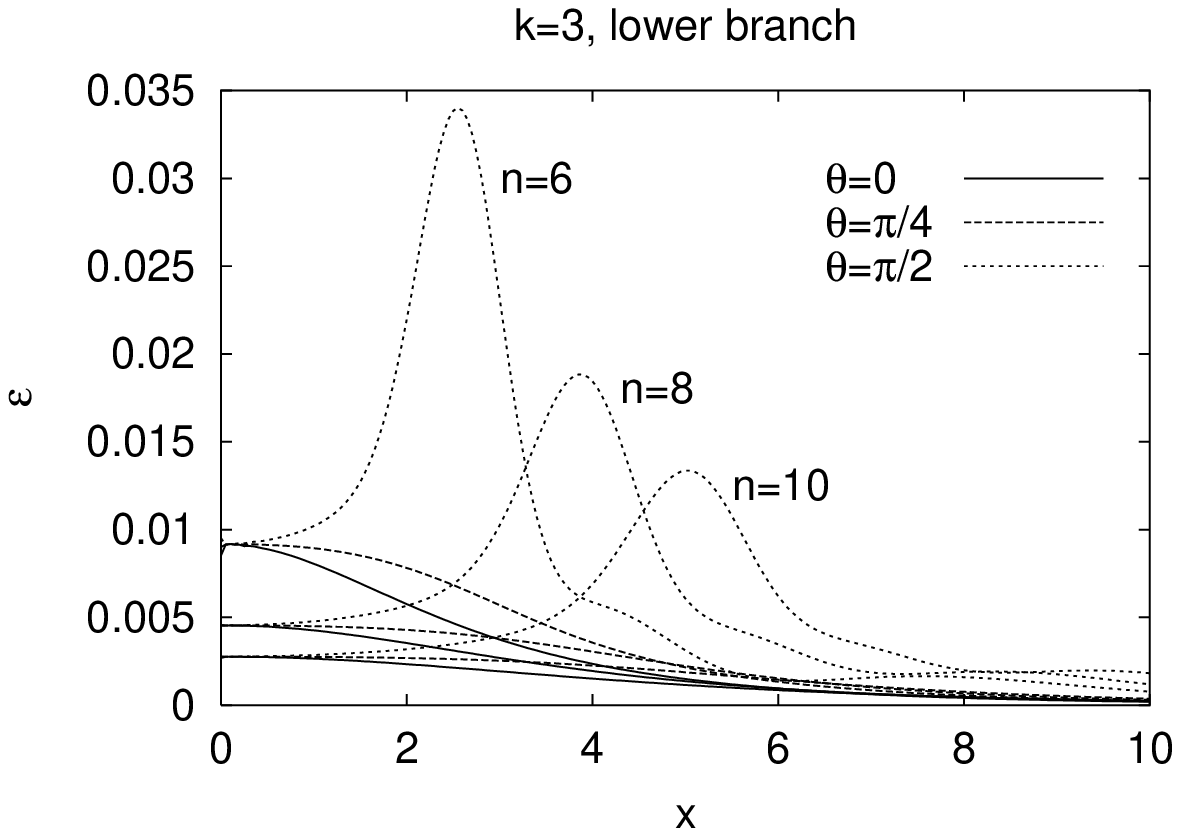}
(b) 
\epsfxsize=6.cm\epsffile{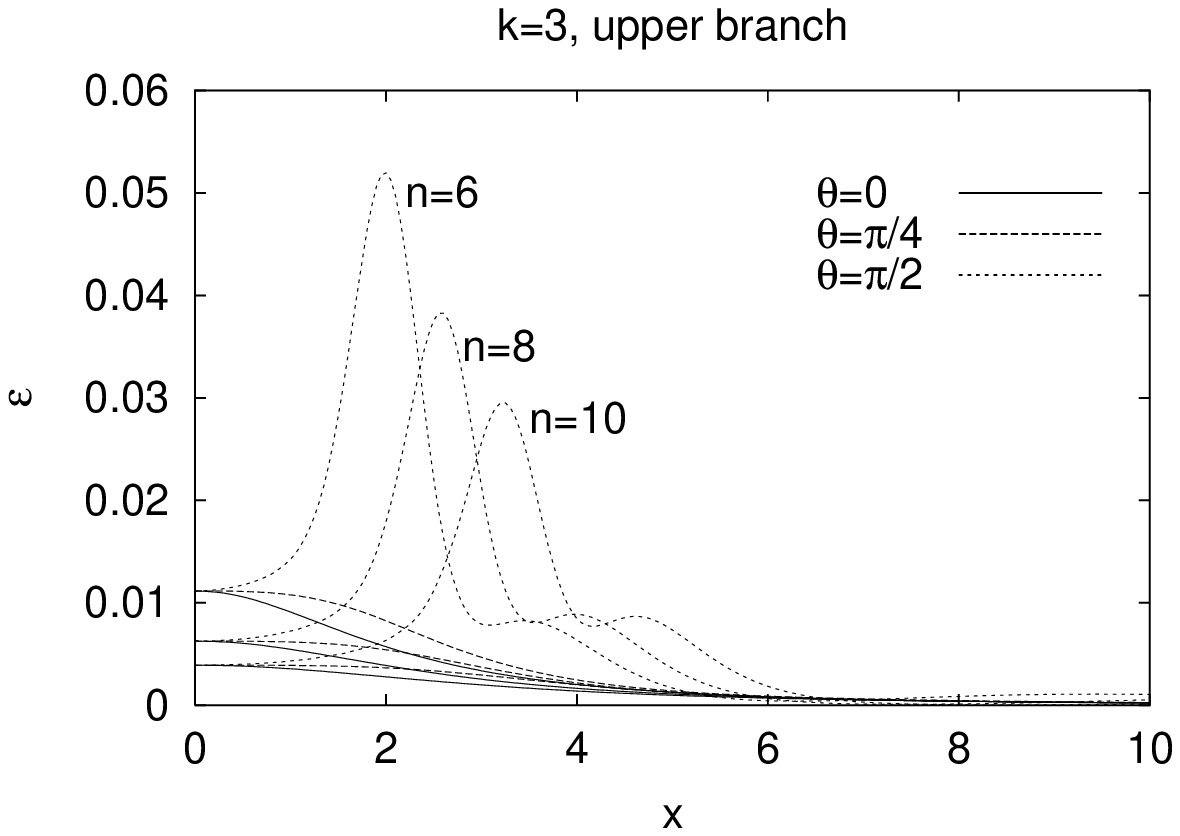}\vspace{0.3cm}
}
{\bf Fig.~6}
{\small
The energy density of the $(3,n)$ EYM solution on the lower mass branch
(a) and on the upper mass branch (b) for $n=6,8,10$.
}\vspace{0.7cm}}

\noindent
\parbox{\textwidth}{
\centerline{(a) 
\epsfxsize=6.cm\epsffile{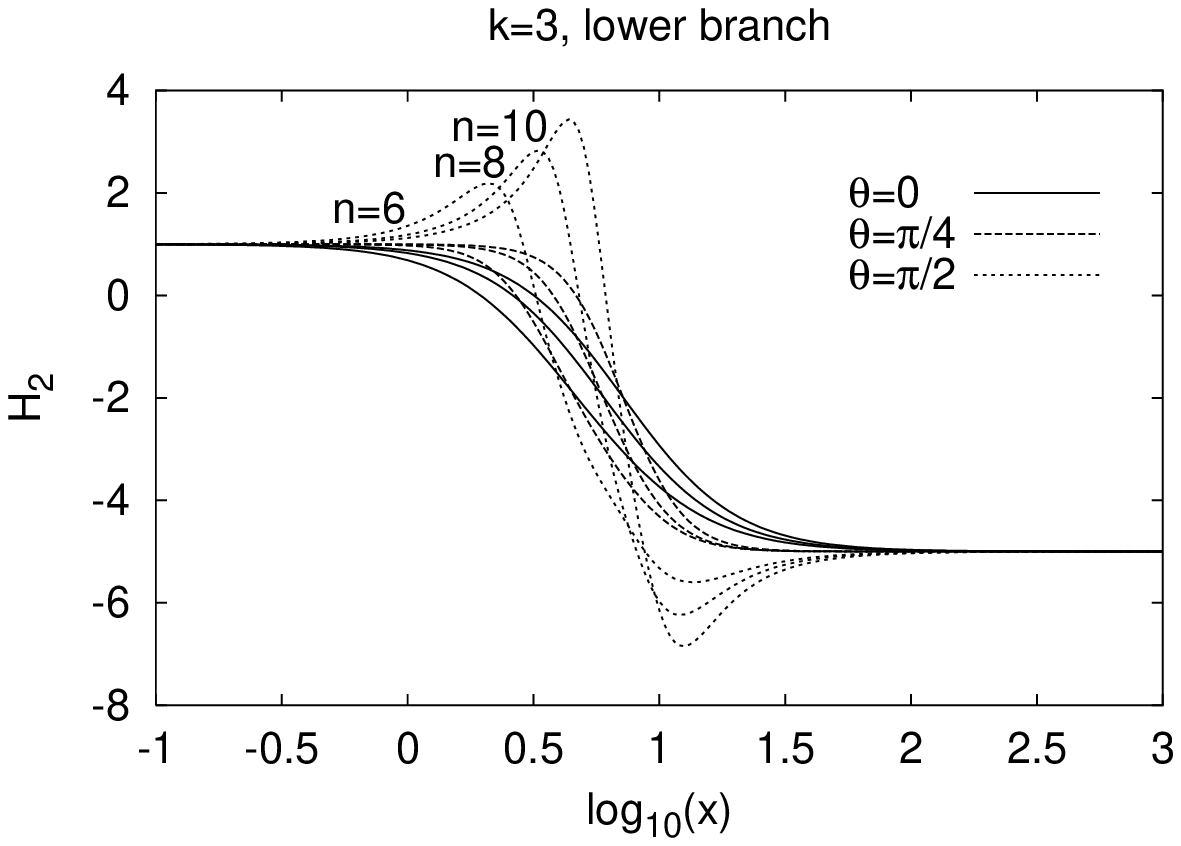}
(b) 
\epsfxsize=6.cm\epsffile{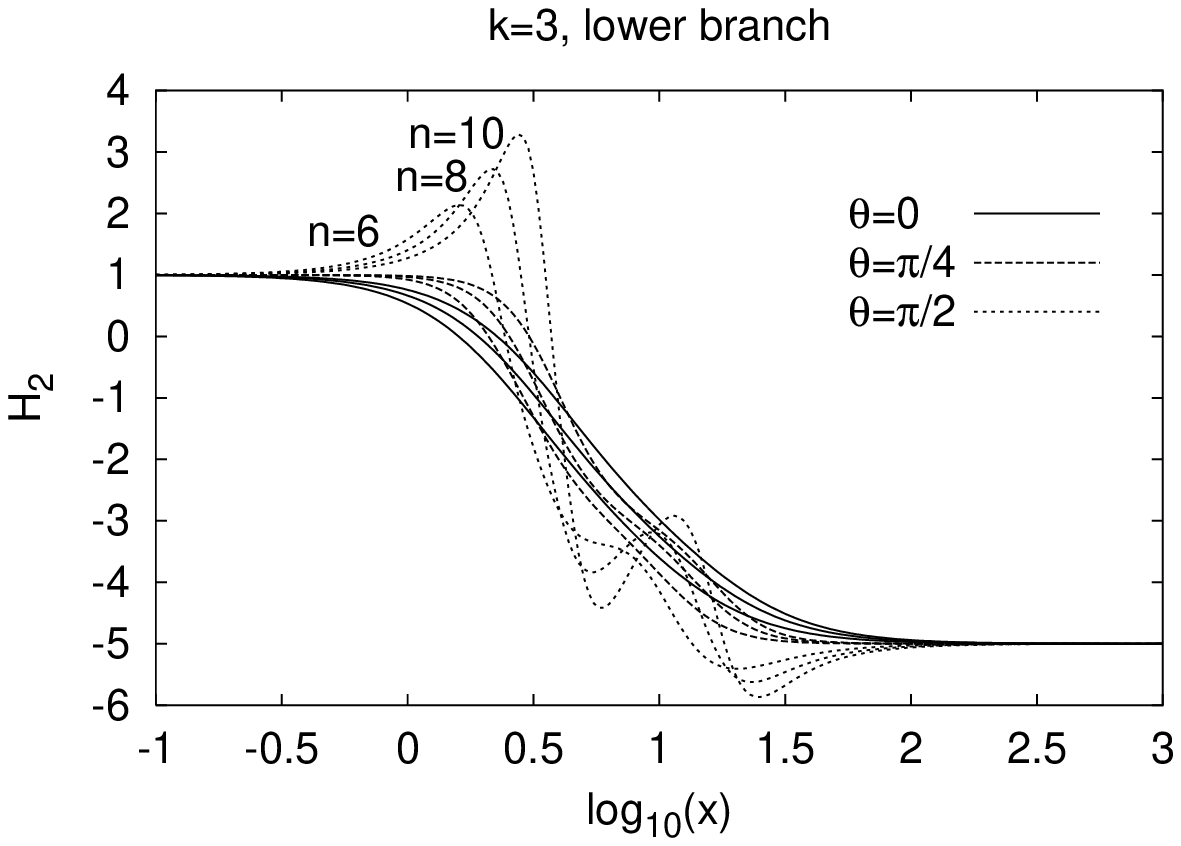}\vspace{0.3cm}
}
{\bf Fig.~7}
{\small
The gauge field function $H_2$ of the $(3,n)$ EYM solutions
on the lower mass branch (a) 
and on the upper mass branch (b) for $n=6,8,10$.
}\vspace{0.7cm}}

\section{Conclusions}

We have constructed numerically new static
regular solutions of EYM theory with axial symmetry.
These solutions are characterized by the integers $(k,n)$,
related to the azimuthal and polar angles, respectively.
In particular, we have obtained solutions
for $k=2$, $n=4-10$, and $k=3$, $n=6-10$.

Like the previously known $(1,n)$ solutions \cite{kk},
the $(2,n)$ and $(3,n)$ solutions 
on the lower branch have a torus-like shape.
On the upper branch of the $(3,n)$ solutions
double tori appear for larger $n$.
The $(1,n)$ solutions (with one node) most likely exist for any
integer $n\ge 1$ and form a single branch.
The new $(2,n)$ and $(3,n)$ solutions have lower bounds on $n$, $n=2k$,
which imply the existence of two branches of solutions
for both values of $k$.
We expect, that the $(2,n)$ and $(3,n)$ solutions
represent only the first sequences of new solutions,
and conjecture the existence of $(k,n)$ solutions
also for higher values of $k$.

The $k=1$ BM solutions are unstable \cite{stab},
and there is all reason to believe, that the new $(k,n)$
solutions are also unstable. 
Having constructed the regular axially symmetric solutions
it appears straightforward
to construct analogous black hole solutions,
and, in particular, to look for rotating solutions \cite{rot}.

{\sl Acknowledgement}

R.I. gratefully acknowledges support by the DAAD,
and B.K. support by the DFG.

\vfill\eject

\end{document}